\documentclass[fp,twocolumn]{jpsj3}
\usepackage{txfonts}
\usepackage{float}
\usepackage{graphicx}
\usepackage{color}
\usepackage{dcolumn}
\usepackage{bm}

\title{Fermi-liquid versus non-Fermi-liquid/'strange-metal' fits to the electrical resistivity\\ in the quantum critical magnetic regime of an unconventional superconductor}

\author{W. Knafo$^{1,2}$\thanks{william.knafo@lncmi.cnrs.fr}, T. Thebault$^{1,3}$, K. Somesh$^1$, G. Lapertot$^4$, G. Knebel$^4$, D. Braithwaite$^4$, and D. Aoki$^2$}
\inst{$^1$ Laboratoire National des Champs Magn\'{e}tiques Intenses - EMFL, CNRS, Univ. Grenoble Alpes, INSA-T, Univ. Toulouse 3, 31400 Toulouse, France\\
$^2$ Institute for Materials Research, Tohoku University, Ikaraki 311-1313, Japan\\
$^3$Institute for Quantum Materials and Technologies, Karlsruhe Institute of Technology, 76131 Karlsruhe, Germany\\
$^4$ Univ. Grenoble Alpes, CEA, Grenoble INP, IRIG, PHELIQS, 38000, Grenoble, France}

\abst{The question of a possible quantum critical point lying inside of a superconducting phase is central for understanding unconventional superconductivity. In various unconventional superconductors, non-Fermi-liquid/'strange-metal' $T^{n}$ variations, with $n<2$, of the electrical resistivity have been identified as the signature of magnetic quantum criticality. However, a difficulty is to prove experimentally that a non-Fermi-liquid/'strange-metal' law identified at temperatures above the superconducting temperature is the signature of an intrinsic zero-temperature quantum critical regime. In the heavy-fermion paramagnet UTe$_2$, unconventional superconductivity develops in the vicinity of a metamagnetic quantum phase transition induced by a magnetic field, and the quantum critical magnetic properties are suspected to play a role for the superconducting mechanism. In this work, we present a comparative analysis of electrical resistivity data collected on two UTe$_2$ samples of different qualities, in magnetic fields tilted by angles $\theta\simeq35-40$~$^\circ$ from $\mathbf{b}$ to $\mathbf{c}$. Fits to the data have been performed either with a Fermi-liquid function $\rho=\rho_0+AT^{2}$ or with a non-Fermi-liquid/'strange-metal' function $\rho=\rho_0+A_nT^n$. Near to a superconducting phase induced beyond 40~T, non-physical residual resistivities $\rho_0<0$ are extracted from the $T^n$ fits, revealing that a 'hidden' Fermi-liquid $T^2$ regime may be ultimately recovered at low temperature. The results obtained here highlight the importance to investigate high-quality samples with low residual resistivity to confirm - or not - the presence of a suspected 'hidden' quantum critical behavior masked by superconductivity.}

\begin{document}
\maketitle

\begin{figure*}[t]
\begin{center}
\includegraphics[width=1\textwidth]{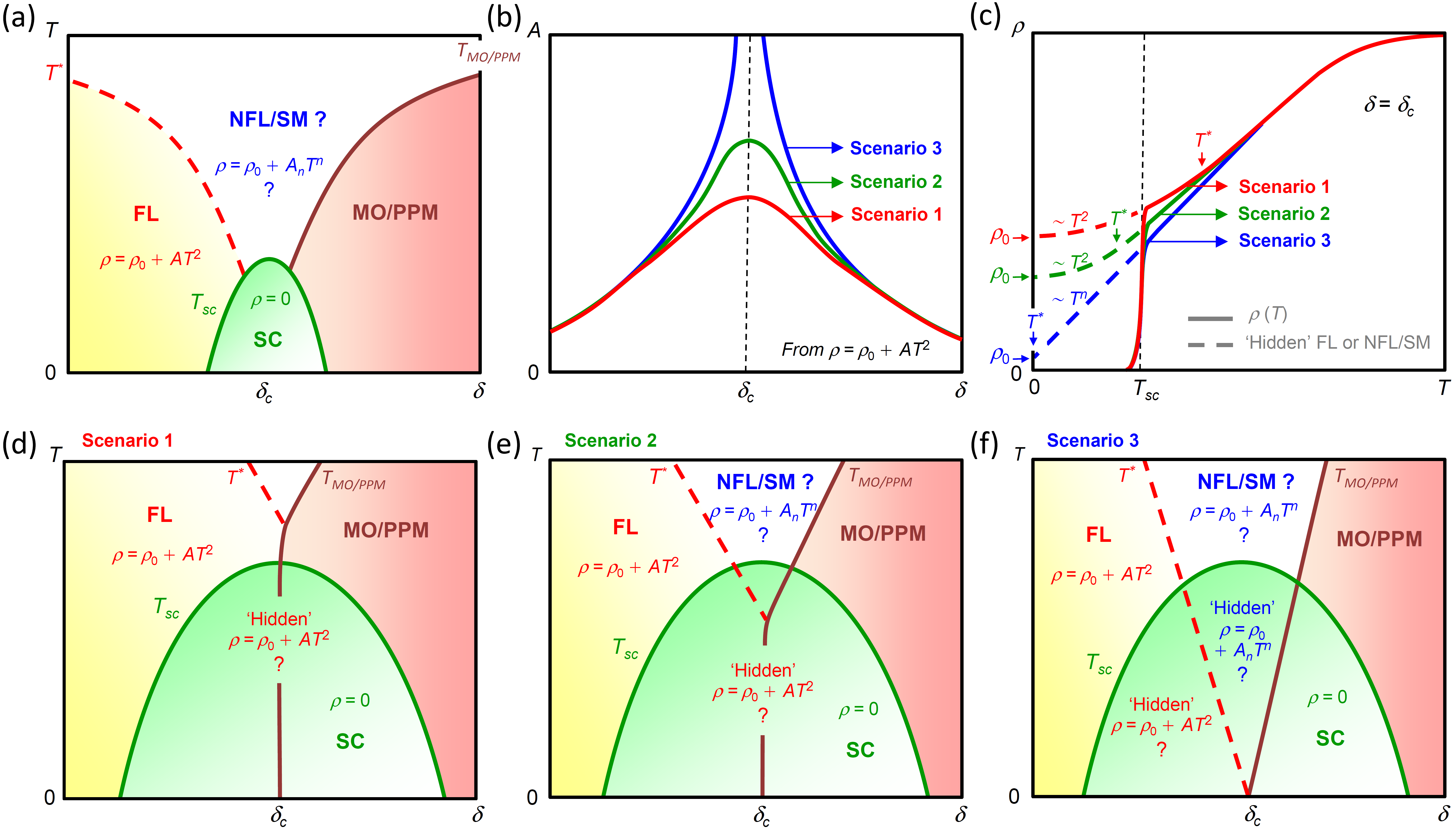}
\caption{\label{Figure1} (a) Quantum critical ($\delta$,$T$) phase diagram, in which $T^*$ delimitates a Fermi-liquid regime, $T_{MO/PPM}$ a magnetically-ordered phase or a polarized paramagnetic regime, and $T_{sc}$ a unconventional superconducting phase. Superconductivity develops in the vicinity of the critical tuning $\delta_c$, at which a quantum phase transition between the Fermi-liquid regime and magnetically-ordered phase or a polarized paramagnetic regime is established. The electrical resistivity varies as $\rho=\rho_0+AT^{2}$ in the Fermi-liquid regime, and as $\rho=\rho_0+A_nT^n$ with $n<2$ in the non-Fermi-liquid/'strange metal' regime sometimes observed near to $\delta_c$ at temperatures $T>T^*,T_{MO/PPM}$. (b) Variation of the Fermi-liquid coefficient $A$ with the tuning parameter $\delta$ and (c) variation of the electrical resistivity $\rho$ with the temperature $T$ at the critical tuning $\delta_c$, for the scenarios 1, 2 and 3 described below. Zooms on the quantum critical ($\delta$,$T$) phase diagrams near to $\delta_c$ for (d) the scenario 1 with $T^*>T_{sc}$ and $T^*$ remaining finite near to $\delta_c$, (e) the scenario 2 with $T^*<T_{sc}$ and $T^*$ remaining finite near to $\delta_c$, and (f) the scenario 3  with $T^*<T_{sc}$ and $T^*$ go to zero at $\delta_c$. FL denotes the Fermi liquid regime, MO/PPM the magnetically-ordered phase or a polarized paramagnetic regime, SC the superconducting phase, NFL/SM the non-Fermi-liquid/'strange-metal' regime.}
\end{center}
\end{figure*}

In correlated-electron systems, a Fermi-liquid regime associated with an effective mass $m^*$ is often observed at temperatures $T$ smaller than a characteristic temperature $T^*$ \cite{Wilson1975,Lee1986,Kadowaki1986,Behnia2004,Knafo2021c}. The electrical resistivity varies then as $\rho=\rho_0+AT^{2}$, where $A\sim m^{*2}$ and $\rho_0$ is the residual resistivity. In the so-called heavy-fermion compounds, heavy effective masses $m^*$ are driven by low-energy magnetic fluctuations \cite{Edwards1992,Rossat1988,Kambe1997,Knafo2021c,Shiba1975,Kitaoka1987,Tokunaga2023} and one can easily tune the physical properties using an adjustable parameter $\delta$, which can be the pressure, a chemical doping, or a magnetic field. The study of these textbook systems since more than forty years allowed to explore the signatures of quantum magnetic criticality, i.e., the critical phenomena observed in the vicinity of a quantum magnetic phase transition \cite{Stewart2001,vonLohneysen2007}. Figure \ref{Figure1}(a) shows a ($\delta$,$T$) phase diagram in which the Fermi-liquid temperature $T^*$ decreases with $\delta$ before a quantum magnetic phase transition, i.e., a magnetic phase transition in the limit $T\rightarrow0$, is established at a critical tuning $\delta_c$. For $\delta>\delta_c$, a magnetically-ordered phase (or a polarized paramagnetic regime under magnetic field) sets in at temperatures below the characteristic temperature $T_{MO/PPM}$. An unconventional superconducting phase often develops at temperatures below the transition temperature $T_{sc}$, in the vicinity of a quantum magnetic phase transition of heavy-fermion materials \cite{Pfleiderer2009} [see Figure \ref{Figure1}(a)]. In several compounds, a low-temperature Fermi-liquid regime is observed in the non-superconducting parts of the phase diagram and a maximum of $m^*$ indicates enhanced magnetic fluctuations at the critical value $\delta_c$ \cite{Araki2002,Knafo2010,Knafo2017,Knafo2009}. In other compounds, deviations from a Fermi-liquid behavior, labeled as non-Fermi-liquid behaviors, have been observed in the low-temperature quantum critical regime near $\delta_c$ \cite{Stewart2001}. Instead of a Fermi-liquid $T^2$ variation, non-Fermi-liquid variations $\rho=\rho_0+A_nT^n$, with a coefficient $n<2$, of the electrical resistivity have been reported \cite{Stewart2001,Custers2003}. Non-Fermi-liquid behaviors have been qualitatively described within the framework of second-order quantum magnetic phase transitions \cite{Hertz1976,Millis1993,Moriya1995}. In these models, the temperature scale $T^*$ vanishes and quantum critical magnetic fluctuations diverge, ending in deviations from a Fermi-liquid behavior at a quantum critical point ($\delta_c$,$T\rightarrow0$).

Following the precursor works on heavy-fermion systems \cite{Pfleiderer2009}, unconventional superconductivity was also found in the vicinity of quantum magnetic phase transitions in various materials, including organic \cite{Salameh2009}, cuprate \cite{Armitage2010}, iron-based \cite{Johnston2010,Shibauchi2014} and nickelate \cite{Li2019} compounds. In these systems, the critical magnetic fluctuations have been suspected to play a role for the mechanism of unconventional superconductivity \cite{Monthoux2007,Pfleiderer2009}, and a strong interest has been given to the investigation of the non-Fermi-liquid regime observed near the magnetic instability. In many materials, a quantum critical linear resistivity versus temperature was identified and a planckian-dissipation mechanism was proposed to lead to $\rho\sim m^*T$ in a so-called 'strange-metal' regime extending to $T\rightarrow0$ \cite{Zaanen2004,Davison2014,Patel2019}. However, the electrical resistivity goes to zero and the magnetic excitation spectra become gapped in the superconducting phase \cite{Rossat1991,Metoki1998,Stock2008,Christianson2008,Yu2009,Stockert2011,Duan2021,Raymond2021}. Low-temperature Fermi-liquid and non-Fermi-liquid/'strange-metal' regimes are therefore masked in the superconducting phase. Figure \ref{Figure1}(c) shows that a low-temperature Fermi-liquid regime where $\rho=\rho_0+AT^{2}$ for $T<T^*$ is always followed by a higher-temperature inflexion point in $\rho(T)$, which can be fitted by $\rho=\rho_0+A_nT^n$ with $n<2$ within a narrow range of temperatures larger than $T^*$. An accurate distinction between Fermi-liquid and non-Fermi-liquid/'strange-metal' regimes requires measurements down to very low temperatures and over a significant range (more than one decade), which is generally not possible in the presence of superconductivity. A challenge is to prove experimentally that a non-Fermi-liquid/'strange-metal' regime identified at temperatures $T>T_{sc}$ is an intrinsic quantum critical property, which would extend down to $T\rightarrow0$ - and, thus, would not be ultimately replaced a Fermi-liquid regime - if superconductivity would not appear.

\begin{figure*}[t]
\begin{center}
\includegraphics[width=0.8\textwidth]{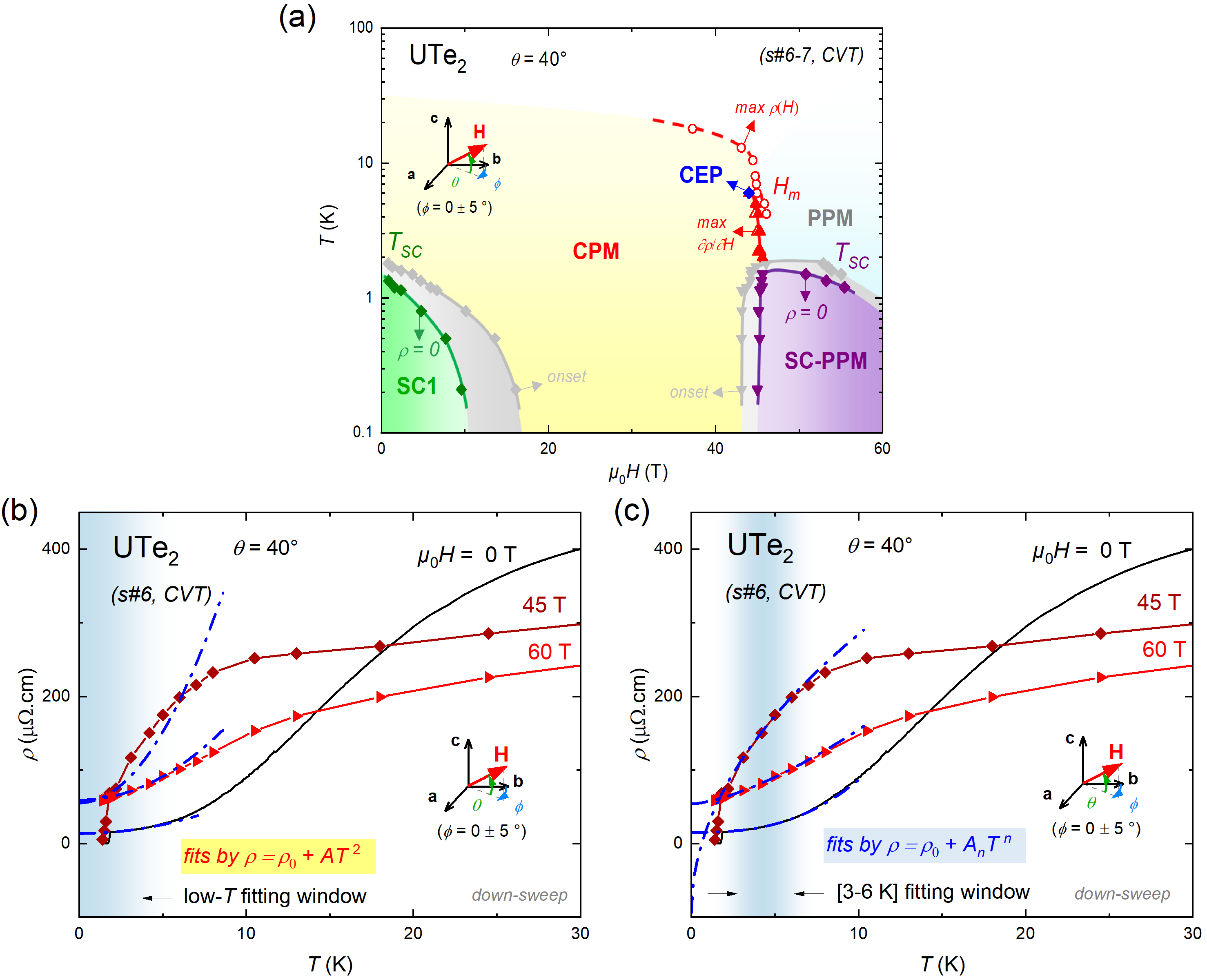}
\caption{\label{Figure2} (a) Magnetic-field-temperature phase diagram of UTe$_2$ in a magnetic field $\mathbf{H}$ tilted by an angle $\theta\simeq40~^\circ$ from $\mathbf{b}$ to $\mathbf{c}$ (from Ref. \cite{Knafo2021a}). Electrical resistivity $\rho$ versus temperature $T$ measured on UTe$_2$ sample $\#6$ at temperatures $T$ from 1.4 to 30~K and magnetic fields $\mu_0H=0$, 45 and 60~T tilted by an angle $\theta\simeq40~^\circ$ from $\mathbf{b}$ to $\mathbf{c}$, together with fits (b) by the Fermi-liquid function $\rho=\rho_0+AT^{2}$ at low temperatures $T>T_{sc}$ and (c) by the function $\rho=\rho_0+A_nT^n$ in the temperature window $3\leq T\leq6$~K (data from Ref. \cite{Knafo2021a}).}
\end{center}
\end{figure*}

In the following, we consider three different scenarios of quantum magnetic criticality [see Figures \ref{Figure1}(b-f)], and we label as 'hidden' the Fermi-liquid and non-Fermi-liquid/'strange-metal' regimes masked by superconductivity, which would be expected at temperatures down to $T\rightarrow0$ without the onset of superconductivity. In the Scenario 1, a Fermi-liquid regime is observed at temperatures $T_{sc}<T<T^*$ for all values of $\delta$, and we can reasonably expect that a 'hidden' Fermi-liquid regime extends virtually at temperatures down to $T\rightarrow0$ [Figures \ref{Figure1}(c-d)]. In this scenario, $T^*$ and $m^*\sim1/T^*$ remain finite at $\delta_c$ [Figures \ref{Figure1}(b,d)]. Scenario 1 permits to describe several heavy-fermion superconductors, for which $T^2$ fits to the electrical resistivity showed a maximum of $A\sim m^{*2}$ at the critical pressure at which superconductivity is also observed \cite{Araki2002,Miyake2008}. Scenario 2 is similar to Scenario 1, but with smaller values of $T^*<T_{sc}$ near to $\delta_c$ leading to a possible variation $\rho=\rho_0+A_nT^n$ with $n<2$ of the electrical resistivity at temperatures $T>T_{sc}$ [Figures \ref{Figure1}(c,e)]. In this scenario, a 'hidden' Fermi-liquid regime is virtually recovered at temperature down to $T\rightarrow0$, and $T^*$ and $m^*\sim1/T^*$ remain finite at $\delta_c$ [Figures \ref{Figure1}(b,c,e)]. Within Scenario 2, electrical-resistivity measurements on an iron-based superconductor showed that, while a 'strange-metal' regime is identified in zero magnetic field at temperatures $T>T_{sc}$, a Fermi-liquid regime is revealed at the lowest temperatures when a magnetic field $H>H_{c,2}$ is applied \cite{Analytis2014}. In the Scenario 3, a variation $\rho=\rho_0+A_nT^n$ with $n<2$ observed at temperatures $T>T_{sc}$ corresponds to a non-Fermi-liquid/'strange-metal' regime,  and a 'hidden' non-Fermi-liquid/'strange-metal' regime extends virtually at temperatures down to $T\rightarrow0$ [Figures \ref{Figure1}(c-f)]. In this scenario, $T^*$ goes to zero at $\delta_c$ and the relation $m^*\sim1/T^*$ is no more valid near $\delta_c$. While the scenarios 1 and 2 are compatible with a first-order character of the quantum magnetic phase transition (see for instance \cite{Knafo2009}), the scenario 3 describes a second-order quantum magnetic phase transition associated with a quantum critical point, where the relevant temperature scales $T^*$ and $T_{MO/PPM}$ both continuously vanish. Scenario 3 fits with the proposition of non-Fermi-liquid/'strange-metal' variations in the electrical resistivity $\rho=\rho_0+AT^n$, with $n<2$ (often $n=1$), identified at temperatures $T>T_{sc}$ and proposed to be a signature of quantum magnetic criticality in heavy-fermion \cite{Mathur1998}, cuprate \cite{Gurvitch1987,Martin1990}, iron-based \cite{Kasahara2010,Sato2025}, nickelate \cite{Zhang2024}, and 'magic-angle' graphene \cite{Cao2020} superconductors. A magnetic field $H$ larger than the superconducting upper magnetic field $H_{c,2}$ was also applied to reveal a 'hidden strange-metal' behavior at temperatures down to $T\rightarrow0$ in cuprate \cite{Dagan2004,Jin2011,Legros1019}, iron-based \cite{Licciardello2019}, organic \cite{Doiron2009}) and nickelate \cite{Hsu2024} superconductors. However, instead of revealing the ground state properties, a magnetic field can sometimes lead to a modification of the magnetic properties, as observed in heavy-fermion systems \cite{Aoki2013,Knafo2021c}. Alternative fits to the electrical resistivity by $\rho=\rho_0+\alpha_1T+\alpha_2T^2$ were also used for a cuprate superconductor and led to the similar conclusion of a linear term $\sim T$ dominating the electrical resistivity at temperatures down to $T\rightarrow0$, near the critical doping \cite{Cooper2009}.

We have recently shown that, in the heavy-fermion paramagnet UTe$_2$ under a magnetic field tilted by an angle $\theta$ from $\mathbf{b}$ to $\mathbf{c}$, a maximum of $A$ extracted using $T^2$ Fermi-liquid fits to the electrical resistivity $\rho$ (measured with a current $\mathbf{I}\parallel\mathbf{a}$) coincides with the domain of stability of a superconducting phase SC-PPM induced in the polarized paramagnetic regime for $\mu_0H\gtrsim\mu_0H_m\simeq40-45$~T and $\theta\simeq35~^\circ$ \cite{Thebault2025}. This maximum of $A$ was identified as the signature of a quantum critical magnetic-fluctuation mode implied in the mechanism of SC-PPM. Using a similar set of electrical-resistivity data on UTe$_2$ (measured with a current $\mathbf{I}\parallel\mathbf{a}$ and magnetic fields tilted from $\mathbf{b}$ to $\mathbf{c}$), Weinberger \textit{et al} also concluded that a quantum critical magnetic regime coincides with the stabilization of SC-PPM \cite{Weinberger2025}. However, they have identified a 'strange-metal' quantum critical behavior from fits by $\rho=\rho_0+A_nT^n$, with $n\simeq1$, to the electrical resistivity at temperatures $T>T_{sc}$. This motivated us to reinvestigate our electrical resistivity data collected on two UTe$_2$ samples $\#6$ and $\#18$, grown by the chemical-vapor-transport and molten-salt-flux techniques, and initially studied in Refs. \cite{Knafo2021a,Thebault2025}, respectively. The different qualities of samples $\#6$ and $\#18$ can be inferred from their residual resistivity ratios $RRR=\rho(300~\rm{K})/\rho(2~\rm{K})$ of 26 and 85, respectively. The re-analysis of our data confirms that the electrical resistivity can be fitted by $\rho=\rho_0+A_nT^n$ with $n\simeq1$ for $T>T_{sc}$ near $H_m$, but negative residual values $\rho_0$ indicate that this law cannot extend down to $T\rightarrow0$. A low-temperature slowing down compatible with the recovery of a $T^2$ Fermi-liquid regime is therefore expected, which finally confirms the pertinence of the $T^2$ Fermi liquid fits performed in \cite{Knafo2021a,Thebault2025}. In the light of our results, we discuss the limits of $T^n$ fits to the electrical resistivity and we emphasize on the importance to investigate samples of the highest quality (with low residual resistivities) to confirm - or not - the suspected presence of a 'strange-metal' quantum critical behavior down to $T\rightarrow0$ in unconventional superconductors.

\begin{figure*}[t]
\begin{center}
\includegraphics[width=0.8\textwidth]{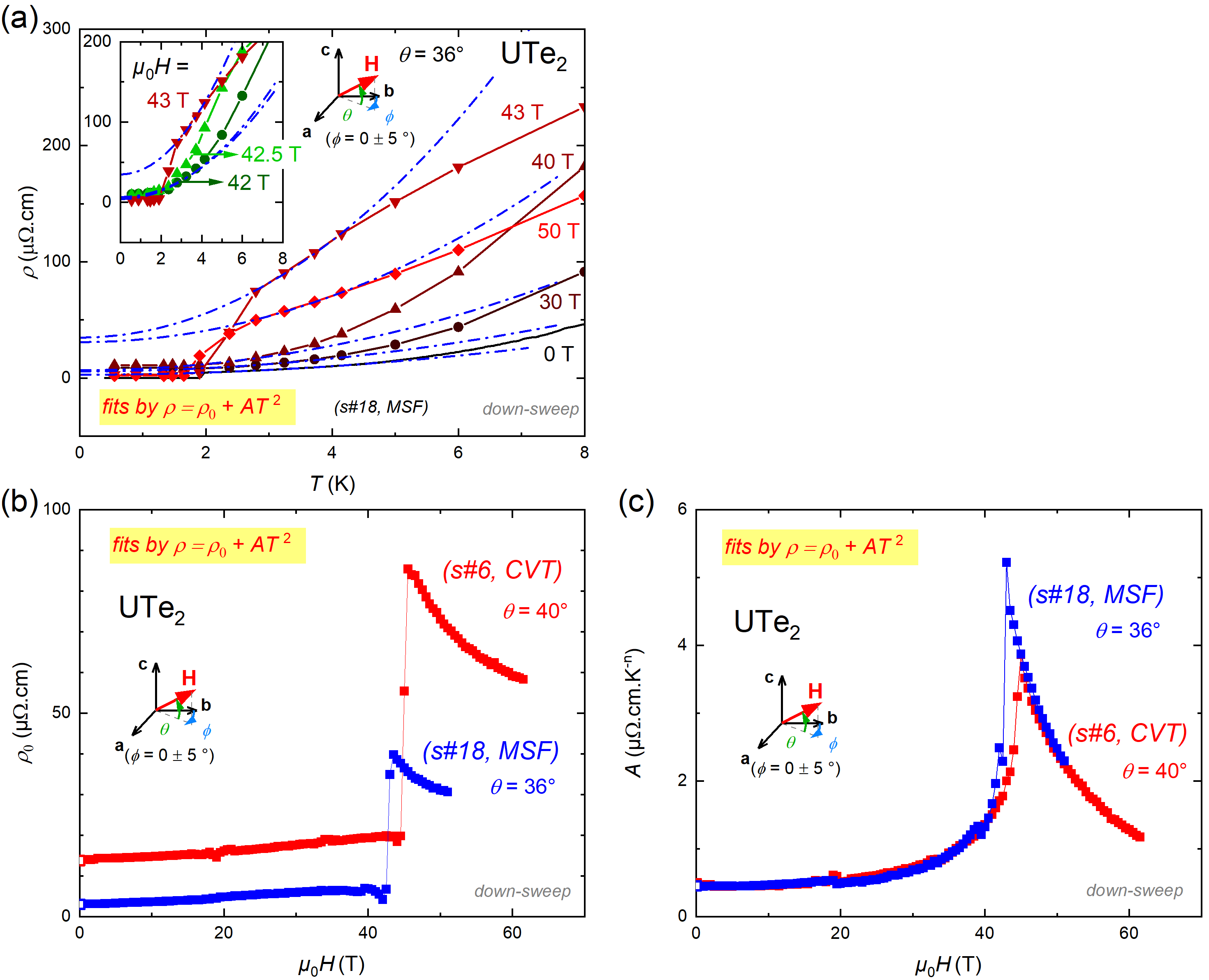}
\caption{\label{Figure3} (a)  Electrical resistivity $\rho$ versus temperature $T$ measured on UTe$_2$ sample $\#18$ at temperatures $T$ from 1.4 to 8~K and magnetic fields $\mu_0H$ from 0 to 50~T tilted by an angle $\theta\simeq6.2~^\circ$ from $\mathbf{b}$ to $\mathbf{c}$, together with fits by the Fermi-liquid function $\rho=\rho_0+AT^{2}$ at low temperatures $T>T_{sc}$. Variations with the magnetic field of the parameters (b) $\rho_0$ and (c) $A$ extracted from the Fermi-liquid fits for samples $\#6$ and $\#18$ with $H$ tilted by angle $\theta\simeq40~^\circ$ and $36~^\circ$, respectively.}
\end{center}
\end{figure*}

Figure \ref{Figure2}(a) presents the magnetic-field-temperature phase diagram of UTe$_2$ obtained in Ref. \cite{Knafo2021a} from electrical resistivity $\rho$ measured with a current $\mathbf{I}\parallel\mathbf{a}$ on samples $\#6$ and $\#7$ issued from the same batch, in a magnetic field $\mathbf{H}$ tilted by an angle $\theta\simeq40~^\circ$ from $\mathbf{b}$ to $\mathbf{c}$ (this angle, initially estimated to $27\pm5~^\circ$, \cite{Knafo2021a} was re-estimated in \cite{Thebault2025}). The phase diagram shows that the low-field superconducting phase SC1 vanishes beyond $\mu_0H_{c2}\simeq10$~T, and that a magnetic-field-induced superconducting phase SC-PPM develops in the polarized paramagnetic regime beyond the metamagnetic field $\mu_0H_m\simeq45$~T, at which the magnetic moments get suddenly polarized \cite{Miyake2021}. Figure \ref{Figure2}(b) shows $\rho$ versus $T$ data measured on sample $\#6$ for a wide range of temperatures $T$ from 1.4 to 30~K and magnetic fields $\mu_0H=0$, 45 and 60~T, together with fits to the low-temperature data by the Fermi-liquid function $\rho=\rho_0+AT^{2}$ for $T>T_{sc}$, from Ref. \cite{Knafo2021a}. The quadratic exponent $A$ is maximum, indicating enhanced critical magnetic fluctuations at the metamagnetic field $\mu_0H_m=45$~T (see also \cite{Knafo2021a}). Figure \ref{Figure2}(c) presents the same set of data together with fits by the function $\rho=\rho_0+A_nT^n$ in the temperature window $3\leq T\leq6$~K, which corresponds to the fitting window and function used by Weinberger \textit{et al} \cite{Weinberger2025}. These fits capture well the change of curvature of $\rho$ versus $T$, which is convex for $\mu_0H=0$ and 60~T and concave for $\mu_0H=45$~T. The temperature at the inflexion point of the electrical resistivity decreases with increasing field, from $\simeq15$~K at $\mu_0H=0$~T to $\simeq2.5$~K at $\mu_0H=45$~T, and then increases up to $\simeq7$~K at $\mu_0H=60$~T. This indicates that the temperature scale of the magnetic fluctuations driving the low-temperature electrical resistivity is minimum at the metamagnetic field $H_m$. In this Figure, one also see that the $T^n$ fit to $\rho$ versus $T$ at $\mu_0H=45$~T leads to a negative value of the residual resistivity $\rho_0$. We will discuss below the magnetic-field variation of the coefficients $A$, $\rho_0$ and $n$ extracted from the fits done on the electrical resistivity of sample $\#6$, but also on the higher-quality sample $\#18$.

\begin{figure*}[t]
\begin{center}
\includegraphics[width=0.8\textwidth]{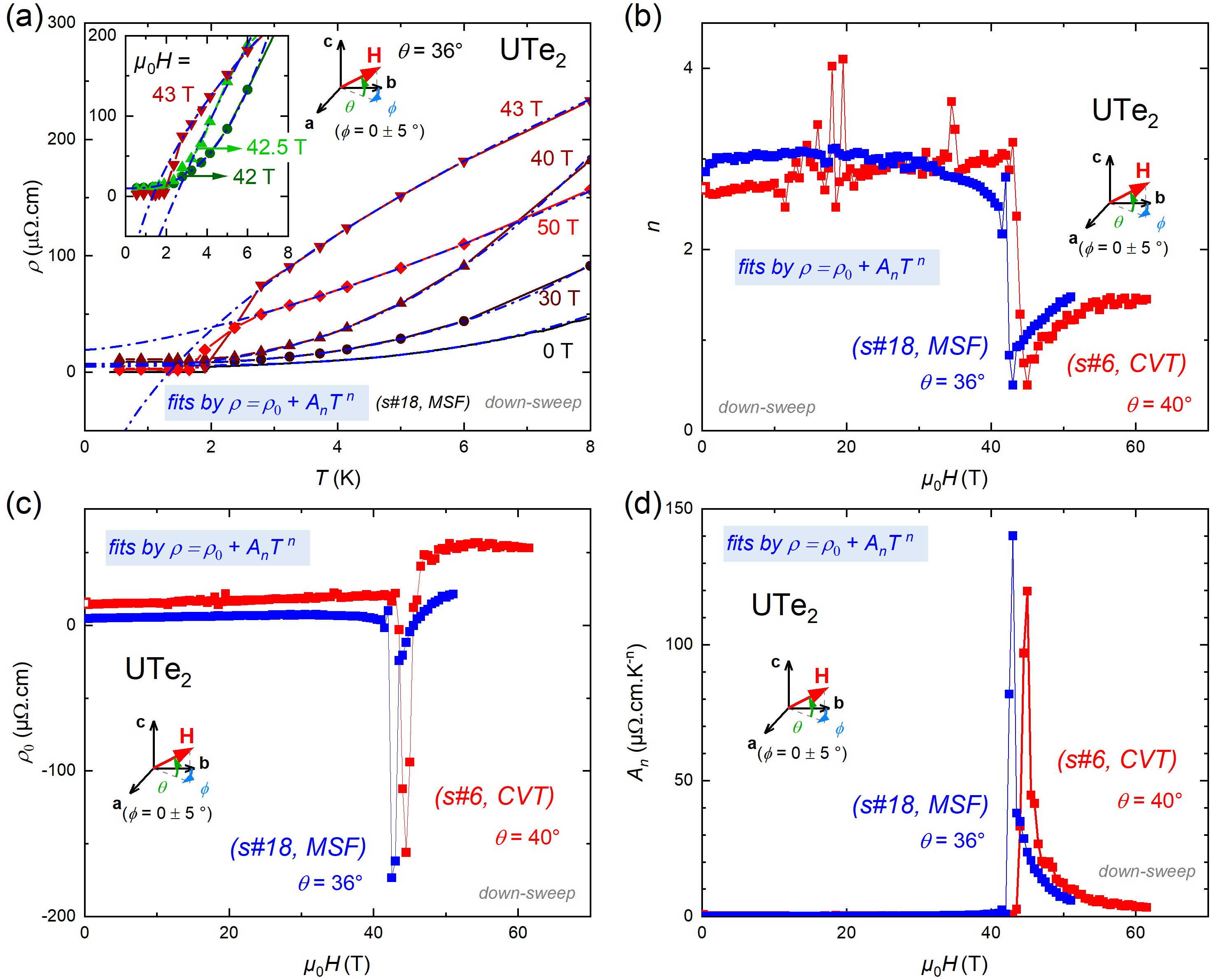}
\caption{\label{Figure4} (a)  Electrical resistivity $\rho$ versus temperature $T$ measured on UTe$_2$ sample $\#18$ at temperatures $T$ from 1.4 to 8~K and magnetic fields $\mu_0H$ from 0 to 50~T tilted by an angle $\theta\simeq6.2~^\circ$ from $\mathbf{b}$ to $\mathbf{c}$, together with fits by $\rho=\rho_0+A_nT^n$ in temperatures $3\leq T\leq6$~K. Variations with the magnetic field of the parameters (b) $n$, (c) $\rho_0$, and (d) $A_n$ extracted from the $T^n$ fits for samples $\#6$ and $\#18$ with $H$ tilted by angle $\theta\simeq40~^\circ$ and $36~^\circ$, respectively.}
\end{center}
\end{figure*}

Figure \ref{Figure3}(a) focuses on Fermi-liquid fits $\rho=\rho_0+AT^{2}$ to the electrical resistivity of sample $\#18$ in a magnetic field $\mu_0\mathbf{H}$ up to 50~T and tilted by an angle $\theta\simeq36~^\circ$ from $\mathbf{b}$ to $\mathbf{c}$, at temperatures up to 8~K, from Ref. \cite{Thebault2025}. Fits have been performed at low temperatures and $T>T_{sc}$. The magnetic-field variations of the residual resistivity $\rho_0$ and the Fermi-liquid parameter $A$ extracted for samples $\#6$ and $\#18$, in magnetic fields $H$ tilted by angles $\theta\simeq40~^\circ$ and $36~^\circ$, respectively, are shown in Figures \ref{Figure3}(b-c). Similar variations of $A$ versus $H$ are found for the two samples, the main difference being a larger enhancement of $A$ near to $\mu_0H_m=43$~T for sample $\#18$, in comparison with that near to $\mu_0H_m=45$~T for sample $\#6$ [Figure \ref{Figure3}(c)]. This is consistent with the previous finding that the maximum of $A$ at $H_m$ is enhanced in the angular window $30\leq\theta\leq35~^\circ$ (see Ref. \cite{Thebault2025}). On the contrary, quite different variations of the residual resistivity $\rho_0$ are found for the two samples [Figure \ref{Figure3}(b)]. At $\mu_0H=0$~T, $\rho_0=13.8$~\textmu$\rm{\Omega cm}$ in sample $\#6$ is more than four times larger than $\rho_0=3$~\textmu$\rm{\Omega cm}$ in sample $\#18$. For both samples, $\rho_0$ slowly increases with $H$ for $H<H_m$, then suddenly increases at $H_m$, before decreasing with $H$ for $H>H_m$. The sudden increase of the residual resistivity $\Delta\rho_0\simeq65$~\textmu$\rm{\Omega cm}$ in sample $\#6$ at $H_m$ is twice larger than $\Delta\rho_0\simeq34$~\textmu$\rm{\Omega cm}$ in sample $\#18$ at $H_m$. While the variations of the coefficient $A$ with the magnetic field are controlled by those of the magnetic fluctuations, which are expected to weakly depend on the sample quality, the variations of the residual resistivity $\rho_0$ with the magnetic field are more sensitive to the sample quality. The residual resistivity, as well as the residual resistivity ratio, are known to be good indicators of sample quality, and the different values of $\rho_0$ observed here are a direct consequence of the different qualities of samples $\#6$ and $\#18$ grown by the chemical-vapor-transport and molten-salt-flux techniques, respectively \cite{Rosa2022,Aoki2024}.

\begin{figure*}[t]
\begin{center}
\includegraphics[width=0.8\textwidth]{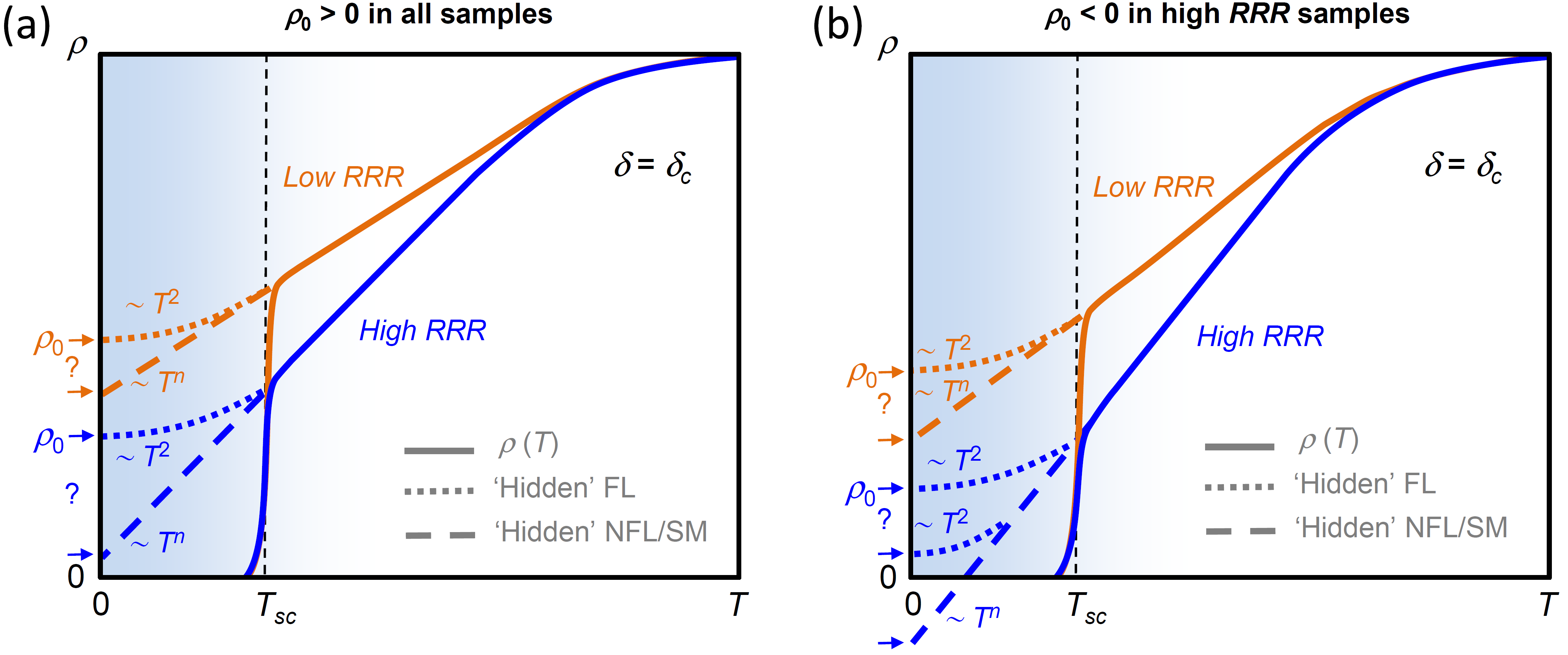}
\caption{\label{Figure5} Electrical resistivity $\rho$ versus temperature $T$ in unconventional superconducting samples with low and high residual-resistivity ratios $RRR$, and possible extrapolations to 'hidden' Fermi-liquid ($\rho=\rho_0+AT^{2}$) or non-Fermi-liquid/'strange-metal' ($\rho=\rho_0+A_nT^n$ with $n<2$) regimes at temperatures $T\rightarrow0$ below the superconducting transition temperature $T_{sc}$. In Panel (a), residual resistivities $\rho>0$ are extrapolated at $T\rightarrow0$ for all samples. In Panel (b), a nonphysical residual resistivity $\rho<0$ is extrapolated on a high-$RRR$ sample at $T\rightarrow0$ from the $T^n$ fit performed for $T>T_{sc}$. FL denotes a Fermi liquid regime and NFL/SM a non-Fermi-liquid/'strange-metal' regime.}
\end{center}
\end{figure*}

Figure \ref{Figure4}(a) presents fits by $\rho=\rho_0+A_nT^n$ to the electrical resistivity of sample $\#18$ in a magnetic field $\mu_0\mathbf{H}$ up to 50~T and tilted by an angle $\theta\simeq36~^\circ$ from $\mathbf{b}$ to $\mathbf{c}$, in temperatures up to 8~K. These fits have been done in the same temperature window $3\leq T\leq6$~K as in Ref. \cite{Weinberger2025}. Contrary to a two-parameter Fermi-liquid fit [see Figure \ref{Figure3}(a)],  this three-parameter fit reproduces well the convex and concave characters of the curves. However, nonphysical negative values of $\rho_0$ are obtained in magnetic fields near to $\mu_0H_m=43$~T. The magnetic-field variations of the exponent $n$, the residual resistivity $\rho_0$ and the coefficient $A_n$ extracted from $T^n$ fits to the electrical resistivity of samples $\#6$ and $\#18$, in magnetic fields $H$ tilted by angles $\theta\simeq40~^\circ$ and $36~^\circ$, respectively, are shown in Figures \ref{Figure4}(b-d). Similar variations of $n$ are found for the two samples, the main difference resulting from the different values of the metamagnetic field $\mu_0H_m=45$~T and 43~T, due to the different angles $\theta=40~^\circ$ and $36~^\circ$, in samples $\#6$ and $\#18$, respectively [Figure \ref{Figure4}(b)]. The exponent $n$ is nearly constant and $\simeq3$ and in magnetic fields $H<H_m$, it reaches a minimum value $\lesssim1$ at $H_m$, before increasing with $H$ for $H>H_m$. These variations of $n$ are similar to that obtained by Weinberger \textit{et al} \cite{Weinberger2025}. Similar variations of the coefficient $A_n$ are also found for samples $\#6$ and $\#18$ [Figure \ref{Figure4}(d)], with $A_n$ being almost constant for $H<H_m$, strongly enhanced at $H_m$ and then decreasing with $H$ for $H>H_m$. For both samples, the residual resistivity $\rho_0$ obtained by $T^n$ fits is almost constant for $H<H_m$, it suddenly decreases to a minimum value at $H_m$, and it increases with $H$ for $H>H_m$ [Figure \ref{Figure4}(c)]. Nonphysical negative values of $\rho_0$ are obtained for the two samples in the vicinity of $H_m$. The values of $\rho_0$ are smaller for sample $\#18$ than for $\#6$, due to the higher crystalline quality of sample $\#18$. Interestingly, the magnetic field window $\mu_0\Delta H\simeq4$~T, where negative values of $\rho_0$ are obtained for the higher quality sample $\#18$, is twice wider than for the lower quality sample $\#6$. This indicates that $T^n$ fits to the electrical resistivity of much-lower-quality UTe$_2$ samples, for which $\rho_0$ is much larger, would lead to positives values of $\rho_0$ at all magnetic fields. However, a nonphysical decrease of $\rho_0$ near to $H_m$ may be identifiable in such lower-quality UTe$_2$ samples. The extraction of non-physical negative values of $\rho_0$ indicates that $T^n$ fits to the electrical resistivity are not relevant in the quantum critical regime of UTe$_2$ under magnetic fields tilted by $\simeq35-40^\circ$ from $\mathbf{b}$ to $\mathbf{c}$. On the contrary, the extracted positive - and apparently reasonable - values of $\rho_0$ [see Figure \ref{Figure3}(b)] indicate that $T^2$ fits are more appropriate here for all field values, even though they could only be done within narrow temperature windows above $T_{sc}$. We note that the $T^n$ fits presented by Weinberger \textit{et al} seem to be compatible with $\rho_0>0$ at magnetic fields up to 60~T and tilted by an angle $\theta=36^\circ$ from $\mathbf{b}$ to $\mathbf{c}$ \cite{Weinberger2025}, which is in contradiction with the results obtained here.

The study of high-quality samples, i.e., samples with a high residual resistivity ratio $RRR$, is therefore crucial to investigate quantum magnetic criticality in the electrical resistivity of unconventional superconductors. Figures \ref{Figure5} illustrates this by considering two general cases.
\begin{itemize}
  \item In Figure \ref{Figure5}(a), the extrapolation of a $T^n$ law observed for $T>T_{sc}$ leads to $\rho_0>0$ in both low-$RRR$ and high-$RRR$ samples. This is compatible with a quantum critical picture, in which the $T^n$ law would continue down to $T\rightarrow0$ if superconductivity was absent. However, one cannot exclude a low-temperature recovery of a $T^2$ Fermi-liquid regime, as identified in the iron-based superconductor BaFe$_2$(As$_{1-x}$P$_x$)$_2$ \cite{Analytis2014}.
  \item In Figure \ref{Figure5}(b), the extrapolation of a $T^n$ law observed for $T>T_{sc}$ leads to $\rho_0>0$ in low-$RRR$ samples and to $\rho_0<0$ in high-$RRR$ samples, as observed here for high-quality UTe$_2$ single crystals. A quantum critical extension of the $T^n$ law down to $T\rightarrow0$ is not relevant, and a low-temperature recovery of a $T^2$ Fermi-liquid regime is expected.
\end{itemize}
A 'hidden' quantum critical behavior continuing at temperature down to $T\rightarrow0$, but masked by superconductivity, was suggested from the observation of $T^n$ variations of the electrical resistivity for $T>T_{sc}$ in a large set of unconventional superconductors \cite{Mathur1998,Gurvitch1987,Martin1990,Kasahara2010,Sato2025,Zhang2024,Cao2020}. Nonphysical negative residual resistivities $\rho_0$ from such $T^n$ laws can also be extracted from measurements done on a few cuprate superconductors \cite{Chien1991,Tyler1997,Yoshida1999,Ando2004}. However, the possibility of a negative residual resistivity extrapolated from a regime identified as a 'strange metal' was scarcely considered (see for instance \cite{Hussey2008}). In the light of the results obtained here, it will be of importance to test the robustness of the propositions for a non-Fermi-liquid/'strange-metal' regime, in different classes of unconventional superconductors, by studying the electrical resistivity of samples of the highest quality, with large residual-resistivity ratios $RRR\gtrsim50-100$ (as in the UTe$_2$ samples studied here).

\begin{acknowledgment}

This work was performed at the Laboratoire National des Champs Magn\'{e}tiques Intenses, a member of the European Magnetic Field Laboratory. We acknowledge financial support from the French National Research Agency collaborative research projects FRESCO No. ANR-20-CE30-0020 and SCATE No. ANR-22-CE30-0040, and from the JSPS KAKENHI Grants Nos. JP22H04933, JP20KK0061, JP24H01641, JP25H01249.

\end{acknowledgment}

\end{document}